\documentclass[10pt]{article}
\usepackage{amsmath}
\usepackage{amssymb}  
\usepackage[T1]{fontenc}
\usepackage{pifont}
\usepackage{pstricks}
\usepackage{amsfonts}
\usepackage{graphicx}
\usepackage{amsmath}
\usepackage{authblk}
\usepackage{pstricks} 
\usepackage{pstricks-add}
\usepackage{caption}
\usepackage{placeins}
\usepackage{pstricks}
\usepackage{pdfpages}
\usepackage{framed}
\usepackage{bm}
\usepackage{epstopdf}
 
\newtheorem{prop}{Proposition}[section]

\def\x2dot{\mathop{x}\limits}
\def\y2dot{\mathop{y}\limits}

\def\bfy2dot{\mathop{\bf y}\limits}
\def\z2dot{\mathop{z}\limits}

\def\csi2dot{\mathop{\xi}\limits}
\def\et2dot{\mathop{\eta}\limits}
\def\bet2dot{\mathop{\beta}\limits}
\def\t2dot{\mathop{\theta}\limits}
\def\s2dot{\mathop{\sigma}\limits}
\def\d2dot{\mathop{\delta}\limits}
\def\q2dot{\mathop{q}\limits}
\def\l2dot{\mathop{\lambda}\limits}
\def\ps2dot{\mathop{{\cal E}}\limits}
\def\tet2dot{\mathop{\theta}\limits}
\def\bfx2dot{\mathop{\bf X}\limits}
\def\bfy2dot{\mathop{\bf y}\limits}
\def\bfq2dot{\mathop{\bf q}\limits}
\def\bbfq2dot{\mathop{\bar {\bf q}}\limits}
\def\w2{\mathop{W}\limits}
\def\xgrande2dot{\mathop{\bf X}\limits}
\def\p02dot{\mathop{P}\limits}
\def\a2dot{\mathop{A}\limits}

\title{Rheonomic systems with nonlinear nonholonomic constraints: the Voronec equations}
\author{Federico Talamucci}
\affil{DIMAI, Dipartimento di Matematica e Informatica ``Ulisse Dini'',\\
Universit\`a degli Studi di Firenze, Italy\\
tel.~+39 055 2751432, fax +39 055 2751452
\\
e-mail: federico.talamucci@unifi.it}
\date{}

\begin{document}
\bibliographystyle{plain}

\setcounter{equation}{0}
\setcounter{ese}{0}
\setcounter{eserc}{0}
\setcounter{teo}{0}               
\setcounter{corol}{0}
\setcounter{propr}{0}

\maketitle

\vspace{.5truecm}





\noindent
{\bf Abstract}. 
One of the earliest formulations of dynamics of nonholonomic systems traces back to 1895 and it is due to 
${\check {\rm C}}$aplygin, who developed his analysis under the assumption that a certain number of the generalized coordinates do not occur neither in the kinematic constraints nor in the Lagrange function. 
A few years later Voronec derived the equations of motion for nonholonomic systems removing the restrictions demanded by the ${\check {\rm C}}$aplygin systems. Although the methods encountered in the following years favour the use of the quasi--coordinates, we will pursue the Voronec method which deals with the generalized coordinates directly.
The aim is to establish a procedure for extending the equations of motion to nonlinear nonholonomic systems, even in the rheonomic case.


\section{Introduction}

\noindent
The context of our analysis is a discrete system of $N$ material points $(P_i,m_i)$, $i=1,\dots, N$. Each point is located with respect to a fixed inertial frame of reference by the triplet ${\bf x}_i\in {\Bbb R}^3$; we enumerate the $3N$ cartesian coordinates in ${\bf X}=({\bf x}_1, \dots, {\bf x}_N)\in {\Bbb R}^{3N}$. The system is possibly exposed to $\ell<3N$ integer (fixed or mobile) independent constraints, so that the preliminary arrangement of the 
problem is settled by the representative vector ${\bf X}(q_1,\dots, q_n,t)$, $n=3N-\ell$, where $(q_1,\dots, q_n)\in {\cal Q}\subseteq {\Bbb R}^n$ are the local lagrangian coordinates.
The relevant feature to join is the presence of nonholonomic constraints formulated by 
$$
\Psi_\nu ({\bf X}, {\dot {\bf X}},t)=0, \quad \nu=1,\dots, k<n
$$
which take the form, owing to the holonomic expression of ${\bf X}$ introduced above:

\begin{equation}
\label{constr}
\Phi_\nu(q_1,\dots, q_n, {\dot q}_1, \dots, {\dot q}_n, t)=0, \quad \nu=1,\dots, k
\end{equation}
The $k$ constraints (\ref{constr}) are assumed to be independent: with no loss in generality, we can assume that it is the jacobian submatrix $J_{({\dot q}_{m+1},\dots, {\dot q}_n)}(\Phi_1, \dots, \Phi_k)$ to be nonsingular, where $m=n-k$, so that they can be written in the form

\begin{equation}
\label{constrexpl}
\begin{cases}
{\dot q}_{m+1}=\alpha_1(q_1, \dots, q_n, {\dot q}_1, \dots, {\dot q}_m, t) \\
\dots  \\
{\dot q}_{m+k}=\alpha_k(q_1, \dots, q_n, {\dot q}_1, \dots, {\dot q}_m, t)
\end{cases}
\end{equation}
with $m=n-k$.

\noindent
If ${\bf X}^{(\mathsf{M})}=(m_1 {\bf x}_1, \dots, m_N {\bf x}_N)\in {\Bbb R}^{3N}$ lists the coordinates of the static moment $(m_1P_1, \dots, m_N P_N)$, the dynamics of the system is summerized by the Newton's equation 

\begin{equation}
\label{ne}
{\dot {\mathcal Q}}={\mathcal F}+{\mathcal R}
\end{equation}
where ${\cal Q}={\dot {\bf X}}^{(\mathsf{M})}$ is the $3N$--vector of linear momenta and ${\mathcal F}=({\mathcal F}_1, \dots, {\mathcal F}_N)$, ${\mathcal R}=({\mathcal R}_1,\dots,{\mathcal R}_N)$ list in ${\Bbb R}^{3N}$ the active forces ${\mathcal F}_i$ and the constraint forces  ${\mathcal R}_i$ concerning each $P_i$, $i=1,\dots, N$.
Equations (\ref{ne}) joined to (\ref{constrexpl}) form a system of $3N+k$ equations for the $3N+n$ unkonwn functions 
$(q_1, \dots, q_n)$ and ${\cal R}$. Actually, up till now no conjecture on the dynamical behaviour of the constraint forces 
has been formulated.

\noindent
The next intent is to extend from the range of holonomic systems the concept of ideal constraint: for those systems 
the constraints are said to be ideal if they play merely the role of restricting the configurations of the system, without affecting the permitted movements of it. Owing to the possible mobility of the configuration, the velocity of the system can be expressed as
$$
{\dot {\bf X}}={\widehat {\dot {\bf X}}}+\frac{\partial {\bf X}}{\partial t}
$$
where ${\widehat {\dot {\bf X}}}$ is any velocity 
consistent with the instantaneous configuration of the system (i.~e.~at a blocked time $t$) and the second term takes into account the mobility. It is suitable to evaluate the properties of ${\cal R}$ (ideal or not) by involving only the term ${\widehat {\dot {\bf X}}}$, which expresses the feasible movements of the systems once the configuration space has been blocked at a certain time. Thus, the set of constraint forces is said to be ideal if 
\begin{equation}
\label{ideal}
{\cal R}\cdot {\widehat {\dot {\bf X}}}=0 \qquad \textrm{for all possible}\;\;{\widehat {\dot {\bf X}}}.
\end{equation}
Definition (\ref{ideal}) entails that the virtual (i.~e.~consistent with the freezed configuration) work of ideal constraints is null.

\noindent
As is known, in the case of entirely holonomic systems, i.~e.~whenever conditions (\ref{constrexpl}) are set aside, we have simply ${\widehat {\dot {\bf X}}}=\sum\limits_{i=1}^n  {\dot q}_i \dfrac{\partial {\bf X}}{\partial q_i}$ for arbitrary $({\dot q}_1, \dots, {\dot q}_n)$, so that ${\cal R}$ is an ideal set if and only if it belongs to the linear space complementary to the tangent space  ${\mathcal T}_{\bf X}$ generated by the $n$ vectors $\frac{\partial {\bf X}}{\partial q_i}$, $i=1,\dots, n$. Equivalently, if we call
\begin{equation}
\label{complagr}
{\mathcal W}^{(q_i)}={\mathcal W}\cdot \dfrac{\partial {\bf X}}{\partial q_i} 
\end{equation}
the $i$--th lagrangian component of any vector ${\mathcal W}\in {\Bbb R}^n$, ${\cal R}$ is ideal if and only if ${\cal R}^{(q_i)}=0$ for any $i=1,\dots, n$.

\noindent
The same scheme is easily extendable to the case of kinematical linear constraints: indeed, whenever (\ref{constrexpl}) are of the form 
\begin{equation}
\label{constrlin}
{\dot q}_{m+\nu}=\sum\limits_{r=1}^m \alpha_{\nu,r}(q_1,\dots, q_n,t){\dot q}_r, \quad \nu=1,\dots, k
\end{equation}
one has
\begin{equation}
\label{subspace}
{\widehat {\dot {\bf X}}}=
\sum\limits_{r=1}^m  {\dot q}_r\left(
\dfrac{\partial}{\partial q_r}{\bf X}(q_1,\dots, q_n,t)+
\sum\limits_{\nu=1}^{k}\alpha_{\nu,r}(q_1,\dots, q_n,t)\dfrac{\partial}{\partial q_{m+\nu}}{\bf X}(q_1,\dots, q_n,t)\right)
\end{equation}
and the arbitrariness of $({\dot q}_1, \dots, {\dot q}_m)$ makes the set of possible displacements, at any blocked time $t$, the linear subspace of ${\mathcal T}_{\bf X}$ generated by the $m$ vectors of (\ref{subspace})  (for $=1,\dots,m$) in brackets. 
Hence, the natural extension of ideal constraint in nonholonomic linear systems demands that the constraint forces have no lagrangian components on that subspace, that is (see also (\ref{complagr}))
\begin{equation}
\label{vincanolid}
{\mathcal R}^{(q_i)}+\sum\limits_{\nu=1}^{k}\alpha_{\nu,i}{\mathcal R}^{(q_{m+\nu})}=0, \qquad i=1,\dots, m.
\end{equation} 
Once again, we see that (\ref{vincanolid}) entails $\sum\limits_{i=1}^n{\mathcal R}^{(q_i)}{\dot q}_i=0$, so that no energy is dissipated.

\noindent
With regard to the general case of nonlinear constraints (\ref{constrexpl}), we take advantage of the linearity of the velocity ${\dot {\bf X}}$ with respect to the generalized velocities ${\dot q}_i$: this is a well--known property in the lagrangian formalism.
We base our next step on remarking that 
the quantity in brackets in (\ref{subspace}) is equal to $\dfrac{\partial {\dot {\bf X}}}{\partial {\dot q}_r}$ for each $r=1,\dots, m$: by considering all the vectors $\sum\limits_{r=1}^m  {\dot q}_r \dfrac{\partial {\dot {\bf X}}}{\partial {\dot q}_r}$ we can attain, for arbitrary $({\dot q}_1, \dots, {\dot q}_m)$, 
the same set (\ref{subspace}). Therefore, in the nonlinear case (\ref{constrexpl}), where the expression 
(analogous to (\ref{subspace})) ${\widehat {\dot {\bf X}}}=
\sum\limits_{r=1}^m  {\dot q}_r \dfrac{\partial {\bf X}}{\partial q_r}+
\sum\limits_{\nu=1}^{k}\alpha_\nu \dfrac{\partial {\bf X}}{\partial q_{m+\nu}}$ cannot provide an ``usable'' space of possible displacements (on varying ${\dot q}_1$, $\dots$, ${\dot q}_m$), 
one finds the access point to specific directions by writing 

\begin{eqnarray}
\label{subspace2}
{\widehat {\dot {\bf X}}}&=&\sum\limits_{r=1}^m  {\dot q}_r \dfrac{\partial {\dot {\bf X}}}{\partial {\dot q}_r}
\\
\nonumber
&=&
\sum\limits_{r=1}^m  {\dot q}_r\left(
\dfrac{\partial}{\partial q_r}{\bf X}(q_1,\dots, q_n,t)+
\sum\limits_{\nu=1}^{k}\dfrac{\partial \alpha_\nu}{\partial {\dot q}_r}(q_1,\dots, q_n,  {\dot q}_1, \dots, {\dot q}_m)\dfrac{\partial}{\partial q_{m+\nu}}{\bf X}(q_1,\dots, q_n,t)\right)
\end{eqnarray}
For arbitrary  $m$--uples $({\dot q}_1, \dots, {\dot q}_m)$ the set (\ref{subspace2}) provides the totality of the possible velocities consistent with the constraints at each position $(q_1, \dots, q_n)$ and, when the configuration space is freezed at a time $t$.
In this way, the constraints forces can be tested with the $m$ vectors appearing in (\ref{subspace2}): 
we are motivated to state that the constraint forces ${\cal R}$ are ideal if they have no components with respect to the $m$ the vectors in brackets in (\ref{subspace2}): 
\begin{equation}
\label{idealnonlin}
{\cal R}^{(q_i)}+\sum\limits_{\nu=1}^{k}\dfrac{\partial \alpha_\nu}{\partial {\dot q}_i}{\cal R}^{(q_{m+\nu})}=0, \qquad \textrm{for each}\;\;i=1,\dots, m.
\end{equation}
As in the linear case, assumption (\ref{idealnonlin}) entails ${\cal R}\cdot {\hat {\dot {\bf X}}}=0$, that is the power of the constraint forces is zero, with respect to all the possible displacements compatible with any blocked configuration of the system.
Finally, it is evident that (\ref{subspace2}) reproduces (\ref{subspace}) whenever (\ref{constrexpl}) are the linear conditions (\ref{constrlin}).

\section{Possible displacements and Newtownian equations}

\noindent
We are going now to develop he calculation of the Newtonian equations (\ref{ne}) along any possible virtual velocity 
${\widehat {\dot {\bf X}}}$, once the set (\ref{subspace2}) has been specified. 
Namely, we will explicit $\left( {\dot {\mathcal Q}}-{\mathcal F}\right)\cdot {\widehat {\dot {\bf X}}}=0$ (see (\ref{ideal}))
for arbitrary ${\dot q}_1$, $\dots$, ${\dot q}_m$ in (\ref{subspace2}). First of all, we calculate 
\begin{eqnarray}
\label{calq}
{\cal Q}=
{\dot {\bf X}^{(\mathsf{M})}}&=&\sum\limits_{r=1}^m\dfrac{\partial {\bf X}^{(\mathsf{M})}}{\partial q_r}
{\dot q}_r+\sum\limits_{\nu=1}^k
\dfrac{\partial {\bf X}^{(\mathsf{M})}}{\partial q_{m+\nu}}\alpha_\nu+\dfrac{\partial {\bf X}^{(\mathsf{M})}}{\partial t}, \\
\nonumber
{\dot {\cal Q}}={\bfx2dot^{..}}^{(\mathsf{M})}&=&
\sum\limits_{r=1}^m \dfrac{\partial {\bf X}^{(\mathsf{M})}}{\partial q_r}{\q2dot^{..}}_r
+\sum\limits_{r,s=1}^m \dfrac{\partial^2 {\bf X}^{(\mathsf{M})}}{\partial q_r \partial q_s}{\dot q}_r {\dot q}_s 
+2\sum\limits_{r=1}^m \sum\limits_{\nu=1}^{k}\dfrac{\partial^2 {\bf X}^{(\mathsf{M})}}{\partial q_r \partial q_{m+\nu}}{\dot q}_r\alpha_\nu
+\sum\limits_{\nu,\mu=1}^k \dfrac{\partial^2 {\bf X}^{(\mathsf{M})}}{\partial q_{m+\nu} \partial q_{m+\mu}}\alpha_\nu \alpha_\mu\\
\label{calqdot}
&+&\sum\limits_{\nu=1}^k\dfrac{\partial {\bf X}^{(\mathsf{M})}}{\partial q_{m+\nu}}\left(
\sum\limits_{r=1}^m \left(\dfrac{\partial \alpha_\nu}{\partial q_r}{\dot q}_r+
\dfrac{\partial \alpha_\nu}{\partial {\dot q}_r} {\q2dot^{..}}_r\right)+ 
\sum\limits_{\mu=1}^k \dfrac{\partial \alpha_\nu}{\partial q_{m+\mu}}\alpha_\mu\right)\\
\nonumber
&+& 2\sum\limits_{r=1}^m \dfrac{\partial^2 {\bf X}^{(\mathsf{M})}}{\partial q_r \partial t}{\dot q}_r +
\sum\limits_{\nu=1}^k\left(2 \dfrac{\partial^2 {\bf X}^{(\mathsf{M})}}{\partial q_{m+\nu} \partial t}\alpha_\nu+ 
\dfrac{\partial {\bf X}^{(\mathsf{M})}}{\partial q_{m+\nu}}\dfrac{\partial \alpha_\nu}{\partial t} \right)+
\dfrac{\partial^2 {\bf X}^{(\mathsf{M})}}{\partial t^2}
\end{eqnarray}
and we state the following 

\begin{prop} Let us call

\begin{equation}
\label{gxi}
\begin{array}{ll}
g_{i,j}(q_1,\dots, q_n, t)=\dfrac{\partial {\bf X}^{(\mathsf{M})}}{\partial q_i}\cdot 
\dfrac{\partial {\bf X}}{\partial q_j}, &
\xi_{i,j,k}(q_1, \dots, q_n, t)=\dfrac{\partial^2{\bf X}^{(\mathsf{M})}}{\partial q_i\partial q_j}\cdot \dfrac{\partial {\bf X}}{\partial q_k},\\
\\
\eta_{i,j}(q_1,\dots, q_n, t)= \dfrac{\partial^2 {\bf X}^{(\mathsf{M})}}{\partial q_i \partial t}\cdot \dfrac{\partial {\bf X}}{\partial q_j}, &
\zeta_i(q_1,\dots, q_n, t)=\dfrac{\partial^2 {\bf X}^{(\mathsf{M})}}{\partial t^2}\cdot 
\dfrac{\partial {\bf X}}{\partial q_j}
\end{array} 
\end{equation}	
for each $i,j,k=1,\dots, n$. Then,the equations of motions of the system ${\bf X}({\bf q},t)$ subject to the nonlinear nonfixed kinematical constraints (\ref{constrexpl}) are
\begin{equation}
\label{vnl2}
\sum\limits_{r=1}^m \left( C_i^r {\q2dot^{..}}_\nu+
\sum\limits_{s=1}^m D_i^{r,s} {\dot q}_r{\dot q}_s + E_i^r {\dot q}_r\right) +G_i={\cal F}^{(q_i)}+
\sum\limits_{\nu=1}^k \dfrac{\partial \alpha_\nu}{\partial {\dot q}_i} {\cal F}^{(q_{m+\nu})}, 
\qquad i=1,\dots, m
\end{equation}
together with (\ref{constrexpl}), where the coefficients, depending on $q_1,\dots, q_n, {\dot q}_1, \dots, {\dot q}_m$ and $t$, are defined by 

\begin{eqnarray}
\nonumber
C_i^r&=&\sum\limits_{\nu,\mu=1}^k\left(g_{i,r}+g_{i,m+\nu}\dfrac{\partial \alpha_\nu}{\partial {\dot q}_r}
+g_{m+\nu,r}\dfrac{\partial \alpha_\nu}{\partial {\dot q}_i}+ g_{m+\nu,m+\mu}\dfrac{\partial \alpha_\mu}{\partial {\dot q}_i}\dfrac{\partial \alpha_\nu}{\partial {\dot q}_r}\right), \\
\nonumber
D_i^{r,s}&=&\xi_{r,s,i}+\sum\limits_{\nu=1}^k\xi_{r,s,m+\nu}\dfrac{\partial \alpha_\nu}{\partial {\dot q}_i}, \\
\label{coeff}
\nonumber
E_i^r &=&  \sum\limits_{\nu,\mu=1}^k \left( 2\left(
\xi_{r,m+\nu,i}+\xi_{r,m+\nu,m+\mu} \dfrac{\partial \alpha_\mu}{\partial {\dot q}_i} \right) \alpha_\nu
+\left( g_{m+\nu,i}+ g_{m+\nu,m+\mu}\dfrac{\partial \alpha_\mu}{\partial {\dot q}_i}\right)\dfrac{\partial \alpha_\nu}{\partial q_r}\right)\\
&+& 
\label{coeffcdef}
2 \left(\eta_{r,i} +\sum\limits_{\nu=1}^k 
\dfrac{\partial \alpha_\nu}{\partial {\dot q}_i}\eta_{r,m+\nu}\right), \\
\nonumber
G_i&=&\sum\limits_{\nu,\mu,p=1}^k\left(
\left(\xi_{m+\nu,m+\mu,i}+ \xi_{m+\nu, m+\mu,m+p} \dfrac{\partial \alpha_p}{\partial {\dot q}_i}\right) \alpha_\nu\alpha_\mu
+\left( g_{m+\nu,i}+g_{m+\nu,m+p}\dfrac{\partial \alpha_p}{\partial {\dot q}_i}\right)
\alpha_\mu \dfrac{\partial \alpha_\nu}{\partial q_{m+\mu}}\right)\\
\nonumber
&+& 2\sum\limits_{\nu=1}^k \left( \eta_{m+\nu, i}+\sum\limits_{\mu=1}^k 
\eta_{m+\nu, m+\mu}\dfrac{\partial \alpha_\mu}{\partial {\dot q}_i}\right) \alpha_\nu+
\sum\limits_{\nu=1}^k \left( g_{m+\nu, i}+\sum\limits_{\mu=1}^k 
g_{m+\nu, m+\mu}\dfrac{\partial \alpha_\mu}{\partial {\dot q}_i}\right) \dfrac{\partial \alpha_\nu}{\partial t}\\
\nonumber
&+& \zeta_i+\sum\limits_{\nu=1}^k \dfrac{\partial \alpha_\nu}{\partial {\dot q}_i}\zeta_{m+\nu}.
\end{eqnarray}
\end{prop}

\noindent
{\bf Proof}. 
Since ${\dot q}_1$, $\dots$, ${\dot q}_m$ are arbitrary, the condition $\left( {\dot {\mathcal Q}}-{\mathcal F}\right)\cdot {\widehat {\dot {\bf X}}}=0$ for any ${\widehat {\dot {\bf X}}}$ as in (\ref{subspace2}) are equivalent to the general dynamical equations
\begin{equation}
\label{proiez}
({\dot {\mathcal Q}}-{\mathcal F})\cdot \left(
\dfrac{\partial {\bf X}}{\partial q_i}+
\sum\limits_{\nu=1}^{k}\dfrac{\partial \alpha_\nu}{\partial {\dot q}_i}\dfrac{\partial {\bf X}}{\partial q_{m+\nu}}\right)=0
\end{equation}
 for each $i=1,\dots,m$.
By taking into account the expression of ${\dot {\cal Q}}$ in (\ref{calqdot}) and the definitions (\ref{gxi}) and (\ref{coeff}) we easily get the left side of (\ref{gxi}). On the other hand, 
${\mathcal F}\cdot \left(
\dfrac{\partial {\bf X}}{\partial q_i}+
\sum\limits_{j=1}^{k}\dfrac{\partial \alpha_j}{\partial {\dot q}_i}\dfrac{\partial {\bf X}}{\partial q_{m+j}}
\right)$ is the right side of (\ref{gxi}) simply by (\ref{complagr}). $\quad \square$

\noindent
The $m+k=n$ equations (\ref{vnl2}) and (\ref{constrexpl}) contain the $n$ unknown functions $q_1(t)$, $\dots$, $q_n(t)$.

\noindent
Whenever the kinematical constraints are stationary, that is none of (\ref{constrexpl}) depends explicitly on $t$, we have $\eta_{i,j}=0$, $\zeta_i=0$ for any $i,j=1,\dots, n$. In this way, we see that in (\ref{vnl2}) the coefficients $C_i^r$ and $D_i^{r,s}$ remain the same, while $E_i^r$ and $G_i$ are reduced, losing the terms beyond the first line of their definition in (\ref{coeff}). This is the case studied in \cite{benenti}, where the equations of motion (\ref{vnl2})
are the Gibbs--Appell equations, via the Gauss principle: actually, we take advantage from this remark in order to 
point out that the left side of (\ref{vnl2}) corresponds to the calculus $\dfrac{\partial S}{\partial {\q2dot^{..}}_i}$
arising in the Appell's method (dating back to \cite{appell}), where $S=\frac{1}{2} {\dot {\cal Q}}\cdot \bfx2dot^{..}$ is the acceleration energy (Gibbs--Appell function). Indeed, owing to (\ref{calqdot}) one has, even in the nonstationary case, $\dfrac{\partial S}{\partial {\q2dot^{..}}_i}=
{\dot {\cal Q}}\cdot \dfrac{\partial \bfx2dot^{..}}{\partial {\q2dot^{..}}_i}={\dot {\cal Q}}\cdot
(\frac{\partial {\bf X}}{\partial q_i}+
\sum\limits_{j=1}^{k}\frac{\partial \alpha_j}{\partial {\dot q}_i}\frac{\partial {\bf X}}{\partial q_{m+j}})$.


\section{Lagrangian-type equations of motion}

\noindent
It is no doubt meaningful to present the equations of motion in terms of the kinetic energy, as it usually occurs starting from the simpler cases through the definition of the Lagrangian function.
This will produce the possibility of a direct comparison with the classic equations of ${\check {\rm C}}$apligin and Voronec, which are expressed in terms of the kinetic energy. Making use of the symbols in (\ref{calqdot}), we write 
\begin{equation}
\label{encin}
T(q_1,\dots, q_n, {\dot q}_1,\dots, {\dot q}_n,t)=
\frac{1}{2}{\mathcal Q}\cdot {\dot {\bf X}}
\end{equation}
as the kinetic energy of the system and we recall the well known property 
$$
{\dot {\mathcal Q}}\cdot \dfrac{\partial {\bf X}}{\partial q_i}=\dfrac{d}{dt}\dfrac{\partial T}{\partial {\dot q}_i}-\dfrac{\partial T}{\partial q_i}.
$$
Owing to (\ref{proiez}) the just mentioned property produces
\begin{equation}
\label{voronec0nonlin}
\dfrac{d}{dt}\dfrac{\partial T}{\partial {\dot q}_i}-\dfrac{\partial T}{\partial q_i}+
\sum\limits_{\nu=1}^k \dfrac{\partial \alpha_\nu}{\partial {\dot q}_i}\left(
\dfrac{d}{dt}\dfrac{\partial T}{\partial {\dot q}_{m+\nu}}-\dfrac{\partial T}{\partial q_{m+\nu}}\right)=
{\cal F}^{(q_i)}+\sum\limits_{j=1}^k \dfrac{\partial \alpha_\nu}{\partial {\dot q}_i} {\cal F}^{(q_{m+\nu})}, 
\qquad i=1,\dots, m.
\end{equation}

\noindent
The interrelations (\ref{constrexpl}) induce the definition of the constrained kinetic energy
$T^*$ as
\begin{equation}
\label{trid}
T^*(q_1,\dots, q_n, {\dot q}_1, \dots, {\dot q}_m, t)=T(q_1, \dots, q_n, {\dot q}_1, \dots, {\dot q}_m, \alpha_1(\cdot), \dots, \alpha_k(\cdot), t)
\end{equation}
intending for each $\alpha_j(\cdot)$, $j=1,\dots, k$, the dependence on $q_1,\dots, q_n, {\dot q}_1, \dots, {\dot q}_m$ and $t$ settled by (\ref{constrexpl}). 
We are going to prove the following
\begin{prop}
	In terms of (\ref{trid}), 
	the equations of motion of the system ${\bf X}({\bf q},t)$ subject to the nonlinear kinematical constraints (\ref{constrexpl}) can be cast in the form
	\begin{equation}
	\label{vnl}
	\dfrac{d}{dt}\dfrac{\partial T^*}{\partial {\dot q}_i}-\dfrac{\partial T^*}{\partial q_i}
	-\sum\limits_{\nu=1}^k\dfrac{\partial T^*}{\partial q_{m+\nu}}\dfrac{\partial \alpha_\nu}{\partial {\dot q_i}}
	-\sum\limits_{\nu=1}^k  B_{i}^\nu \dfrac{\partial T}{\partial {\dot q}_{m+\nu}}
	={\cal F}^{(q_i)}+\sum\limits_{\nu=1}^k \dfrac{\partial \alpha_\nu}{\partial {\dot q}_i} {\cal F}^{(q_{m+\nu})}, 
	\qquad i=1,\dots, m
	\end{equation}
	joined ith (\ref{constrexpl}), where we defined 
	\begin{equation}
	\label{b}
	\begin{array}{ll}
	B_{i}^{\nu}(q_1,\dots, q_n, {\dot q}_1, \dots, {\dot q}_m,t)
	&=\sum\limits_{r=1}^m
	\left( \dfrac{\partial^2 \alpha_\nu}{\partial {\dot q}_i \partial q_r}{\dot q}_r +
	\dfrac{\partial^2 \alpha_\nu}{\partial {\dot q}_i \partial {\dot q}_r}{\q2dot^{..}}_r\right)
	-\dfrac{\partial \alpha_\nu}{\partial q_i}\\
	&+\sum\limits_{\mu=1}^k\left( \dfrac{\partial^2 \alpha_\nu}{\partial {\dot q}_i \partial q_{m+\mu}}\alpha_\mu-
	\dfrac{\partial \alpha_\mu}{\partial {\dot q}_i}\dfrac{\partial \alpha_\nu}{\partial q_{m+\mu}}\right)+
	\dfrac{\partial^2 \alpha_\nu}{\partial {\dot q}_i \partial t}.
	\end{array}
	\end{equation}
\end{prop}

\noindent
{\bf Proof}. The following relations

\begin{equation}
\label{r12}
\begin{array}{ll}
\dfrac{\partial T}{\partial {\dot q}_i}=\dfrac{\partial T^*}{\partial {\dot q}_i}-\sum\limits_{\nu=1}^k 
\dfrac{\partial T}{\partial {\dot q}_{m+\nu}}\dfrac{\partial \alpha_\nu}{\partial {\dot q}_i}, & i=1,\dots, m \\
\\
\dfrac{\partial T}{\partial q_i}=\dfrac{\partial T^*}{\partial q_i}-\sum\limits_{\nu=1}^k 
\dfrac{\partial T}{\partial {\dot q}_{m+\nu}}\dfrac{\partial \alpha_\nu}{\partial q_i}, & i=1,\dots, n
\end{array}
\end{equation}
allow us to write (\ref{voronec0nonlin}) in terms of $T^*$: in order to get (\ref{vnl}) it suffices to have in mind that
$$
\begin{array}{l}
-\dfrac{d}{dt}\left(\dfrac{\partial T}{\partial {\dot q}_{m+\nu}}\dfrac{\partial \alpha_\nu}{\partial {\dot q}_i}\right)+
\dfrac{d}{dt}\left(\dfrac{\partial T}{\partial {\dot q}_{m+\nu}}\right)\dfrac{\partial \alpha_\nu}{\partial {\dot q}_i}=
-\dfrac{\partial T}{\partial {\dot q}_{m+\nu}}
\dfrac{d}{dt}\left(\dfrac{\partial \alpha_\nu}{\partial {\dot q}_i}\right)=\\
\\=
-\dfrac{\partial T}{\partial {\dot q}_{m+\nu}}\left(
\sum\limits_{s=1}^m
\left( 
\dfrac{\partial^2 \alpha_\nu}{\partial {\dot q}_i \partial q_s}{\dot q}_s +
\dfrac{\partial^2 \alpha_\nu}{\partial {\dot q}_i \partial {\dot q}_s}{\q2dot^{..}}_s
\right)
+
\sum\limits_{\mu=1}^k
\dfrac{\partial^2 \alpha_\nu}{\partial {\dot q}_i \partial q_{m+\mu}}\alpha_\mu+
\dfrac{\partial^2 \alpha_\nu}{\partial {\dot q}_i \partial t}\right). \qquad \square
\end{array}
$$
The $m$ equations (\ref{vnl}) contain the $n$ unknown functions $q_1(t)$, $\dots$, $q_n(t)$ and they have to be coupled with (\ref{constrexpl}). 
Obviously, in the terms $\dfrac{\partial T}{\partial {\dot q}_{m+\nu}}$ the arguments ${\dot q}_{m+1}$, $\dots$, ${\dot q}_n$ are removed by means of (\ref{constrexpl}).

\noindent
Equations of motion in the form (\ref{vnl}) leave open to more than one comment and comparison with special cases, shown below.

\noindent
Whenever $\alpha_1$, $\dots$, $\alpha_k$  are the linear functions (\ref{constrlin}) we have for any $i,s=1,\dots, m$, $\nu = 1,\dots, k$ 
$$
\dfrac{\partial \alpha_\nu}{\partial {\dot q}_i}=\alpha_{\nu,i}, \quad
\dfrac{\partial^2 \alpha_\nu}{\partial {\dot q}_i \partial {\dot q}_s}=0, \quad
\dfrac{\partial^2 \alpha_\nu}{\partial {\dot q}_i \partial t}=\dfrac{\partial \alpha_{\nu,i}}{\partial t}
$$
and the coefficients take the antisymmetric expressions
$$
\begin{array}{l}
\sum\limits_{r=1}^m\dfrac{\partial^2 \alpha_\nu}{\partial {\dot q}_i \partial q_r}{\dot q}_r-
\dfrac{\partial \alpha_\nu}{\partial q_i}
=\sum\limits_{r=1}^m \left(
\dfrac{\partial \alpha_{\nu,i}}{\partial q_r}-\dfrac{\partial \alpha_{\nu, r}}{\partial q_i}\right){\dot q}_r, \quad
 \\
\\
\sum\limits_{\mu=1}^k\left( \dfrac{\partial^2 \alpha_\nu}{\partial {\dot q}_i \partial q_{m+\mu}}\alpha_\mu-
\dfrac{\partial \alpha_\mu}{\partial {\dot q}_i}\dfrac{\partial \alpha_\nu}{\partial q_{m+\mu}}\right)=
\sum\limits_{\mu=1}^k \sum\limits_{r=1}^m
\left(\dfrac{\partial \alpha_{\nu, i}}{\partial q_{m+\mu}}\alpha_{\mu,r}-
\dfrac{\partial \alpha_{\nu, r}}{\partial q_{m+\mu}}\alpha_{\mu,i}\right){\dot q}_r.
\end{array}
$$
Consequently, the equations of motion (\ref{vnl}) are of the form 
\begin{equation}
\label{voronec}
\dfrac{d}{dt}\dfrac{\partial T^*}{\partial {\dot q}_i}-\dfrac{\partial T^*}{\partial q_i}
-\sum\limits_{\nu=1}^k\alpha_{\nu,i}\dfrac{\partial T^*}{\partial q_{m+\nu}}
-\sum\limits_{\nu=1}^k\sum\limits_{r=1}^m 
\beta_{ir}^\nu{\dot q}_r\dfrac{\partial T}{\partial {\dot q}_{m+\nu}}=
{\mathcal F}^{(q_i)}+\sum\limits_{\nu=1}^{k} \alpha_{\nu,i} {\cal F}^{(q_{m+\nu})}
\quad i=1,\dots, m
\end{equation}
where, for any $r=1,\dots, m$ and $\nu=1,\dots, k$:
\begin{equation}
\label{beta}
\beta_{ir}^\nu(q_1,\dots, q_n, t)=
\dfrac{\partial \alpha_{\nu, i}}{\partial q_r}-
\dfrac{\partial \alpha_{\nu,r}}{\partial q_i}+
\sum\limits_{\mu=1}^k\left(
\dfrac{\partial \alpha_{\nu, i}}{\partial q_{m+\mu}}\alpha_{\mu,r}-
\dfrac{\partial \alpha_{\nu, r}}{\partial q_{m+\mu}}\alpha_{\mu,i}
\right)+\dfrac{\partial \alpha_{\nu,i}}{\partial t}.
\end{equation}
We see that in the linear case the second derivatives of the constraint functions (\ref{constrexpl}) disappear.

\noindent
The stationary case $\alpha_{k, i}(q_1, \dots, q_n)$ for any $k=1,\dots, \nu$ in (\ref{constrlin}) corresponds to the classic Voronec equations formulated in \cite{voronec} and discussed in \cite{neimark}; the only additional term in the non--stationary case (\ref{vnl}) is the last second derivative in (\ref{b}). The same case can be put into the context
of equations of motion in quasi--coordinates: in wider terms, a nonholonomic system with constraints of the form 
$$
\sum\limits_{j=1}^n a_{m+\nu,j}(q_1,\dots, q_n){\dot q}_j=0, \quad \nu=1,\dots, k
$$
can be treated by defining the quasi--coordinates $\pi_1$, $\dots$, $\pi_n$ so that the last ones verify
${\dot \pi}_{m+\nu}=\sum\limits_{j=1}^n a_{m+\nu,j}{\dot q}_j$, $\nu=1,\dots, k$, and the first $m$ are related to the coordinates by linear equations 
${\dot \pi}_i=\sum\limits_{j=1}^n a_{i,j}(q_1,\dots, q_n){\dot q}_j$, $i=1,\dots, m$, so that the matrix $a_{r,s}$, $r,s=1,\dots, n$ is nonsingular. By means of quasi--coordinates the equations of motion derived by Hamel (\cite{hamel}, \cite{neimark}) exhibit the known form
\begin{equation}
\label{hamel}
\dfrac{d}{dt}\dfrac{\partial T^*}{\partial {\dot \pi}_i}-\dfrac{\partial T^*}{\partial \pi_i}
+\sum\limits_{\ell=1}^n\sum\limits_{j=1}^m \gamma_{i,\ell,j}
\dfrac{\partial T^*}{\partial {\dot \pi}_\ell}{\dot \pi}_j=
{\cal F}^{(\pi_i)}
\quad i=1,\dots, m
\end{equation}
where 
$\gamma_{i,\ell,j}=\sum\limits_{r,s=1}^n b_{r,i}b_{s,j}
\left(\dfrac{\partial a_{\ell,r}}{\partial q_s}-
\dfrac{\partial \alpha_{\ell,s}}{\partial q_r}\right)$ for each $\ell=1,\dots, n$ and $i,j=1,\dots, m$ and $b_{\rho,\sigma}$ are the entries of the inverse of the matrix $a_{r,s}$, with $\rho,\sigma,r,s=1,\dots,n$; moreover,   ${\cal F}^{(\pi_i)}=\sum\limits_{s=1}^n{\cal F}^{(q_s)}b_{s,i}$ is the generalized force relative to the $i$--th quasi--coordinate.

\noindent
Keeping the assumption of stationary constraints, we see that (\ref{voronec}) can be traced in (\ref{hamel}), with the special selection ${\dot \pi}_i={\dot q}_i$, $i=1,\dots, m$: actually, the $n\times n$ matrix $a_{r,s}$ is $\left(\begin{array}{cc} {\Bbb I}_m & {\Bbb O}_{m,k} \\ -{\Bbb A} & {\Bbb I}_k \end{array} \right)$, with ${\Bbb I}_m$ (resp.~$k$) is the identy matrix of order $m$ (resp.~$k$), ${\Bbb O}_{m,k}$ is the $m\times k$ null matrix, ${\Bbb A}$ is the $k\times m$ matrix with entries $\alpha_{\nu, r}$ (see (\ref{constrlin})). The inverse matrix $b_{\rho,\sigma}$ is simply $\left(\begin{array}{cc} {\Bbb I}_m & {\Bbb O}_{m,k} \\ {\Bbb A} & {\Bbb I}_k \end{array} \right)$
so that, comparing with (\ref{beta}):
$$
\begin{array}{ll}
\gamma_{i,\ell,j}=0& \textrm{for}\;\;i,\ell,j=1,\dots,m\\
\\
\gamma_{i,m+\nu,j}=\beta_{ij}^\nu & \textrm{for}\;\;i,j=1,\dots, m,\;\;\nu=1,\dots,k
\end{array}
$$
where the time derivative in (\ref{beta}) is zero.
Moreover, owing to the property (see (\cite{neimark})) $\dfrac{\partial}{\partial \pi_i}=\sum\limits_{j=1}^n b_{j,i}\dfrac{\partial}{\partial q_j}$, one has for each $i=1,\dots,m$
$$
\begin{array}{l}
\dfrac{\partial T^*}{\partial \pi_i}=\sum\limits_{j=1}^n \dfrac{\partial T^*}{\partial q_j}b_{j,i}=
\dfrac{\partial T^*}{\partial q_i}+\sum\limits_{\nu=1}^k \alpha_{\nu,i}\dfrac{\partial T^*}{\partial m+\nu}, \\
{\cal F}^{(\pi_i)}={\mathcal F}^{(q_i)}+\sum\limits_{\nu=1}^{k} \alpha_{\nu,i} {\cal F}^{(q_{m+\nu})}
\end{array}
$$
and (\ref{voronec}) is the same as (\ref{hamel}), whenever quasi--coordinates are true coordinates.

\noindent
Even though the task of tracking down suitable quasi--coordinates fit to the purpose of solving the mathematical problem is 
by no means simple, equations of motion are most often presented in the form (\ref{hamel}) rather than (\ref{voronec}).
In addition, the latter one does not demand the calculation of the inverse matrix.

\noindent
The opportunity of disentangling the equations of motion from the constraints demands a special comment:
let us go back for now to the general nonlinear set (\ref{vnl})
As we already said, system (\ref{voronec}) must be coupled with the constraints expressions (\ref{constrexpl}), because of the presence of $q_{m+1}$, $\dots$, $q_n$. In special cases, it may happen that the set of lagrangian coordinates $q_{m+1}$, $\dots$, $q_n$ is not present in $T$, in the coefficients $\alpha_\nu$, $\nu=1, \dots, k$ nor in the generalized forces ${\cal F}^{(q_i)}$ for any $i=1,\dots, m$.
In that case, the coefficients (\ref{b}) for each $i=1,\dots, m$ reduce to 
$$
\begin{array}{l}
B_{i}^{\nu}(q_1,\dots, q_m, {\dot q}_1, \dots, {\dot q}_m,t)
=\sum\limits_{r=1}^m
\left( \dfrac{\partial^2 \alpha_\nu}{\partial {\dot q}_i \partial q_r}{\dot q}_r +
\dfrac{\partial^2 \alpha_\nu}{\partial {\dot q}_i \partial {\dot q}_r}{\q2dot^{..}}_r\right)
-\dfrac{\partial \alpha_\nu}{\partial q_i}+
\dfrac{\partial^2 \alpha_\nu}{\partial {\dot q}_i \partial t},\\
\\
\alpha_\nu(q_1,\dots, q_m, {\dot q}_1, \dots, {\dot q}_m,t), \;\;\nu=1,\dots,k
\end{array}
$$
so that the $m$ equations of system (\ref{vnl}) can be solved independently from (\ref{constrexpl}), 
since $T^*=T^*(q_1,\dots, q_m,$ ${\dot q}_1,$ $\dots, {\dot q}_m,t)$, the same for the forces terms.
The linear case (\ref{constrlin}), where in addition the functions $\alpha_{\nu,r}$ are assumed not to depend on $t$, leads to the known equations formulated by ${\check {\rm C}}$aplygin (\cite{capligin}) some years before the Voronec equations (see also \cite{neimark}): the coefficients come directly from (\ref{beta}), under the stated assumptions:

\begin{equation}
\label{capligin}
\dfrac{d}{dt}\dfrac{\partial T^*}{\partial {\dot q}_i}-\dfrac{\partial T^*}{\partial q_i}
-\sum\limits_{\nu=1}^k\sum\limits_{j=1}^m 
\left(\dfrac{\partial \alpha_{\nu, i}}{\partial q_j}-
\dfrac{\partial \alpha_{\nu,j}}{\partial q_i}\right)
{\dot q}_j\dfrac{\partial T}{\partial {\dot q}_{m+\nu}}=
{\cal F}^{(q_i)}+\sum\limits_{j=1}^{k} \alpha_{j,i} {\cal F}^{(q_{m+j})}
\quad i=1,\dots, m
\end{equation}
Although the Capligin's assumptions may appear demanding, it is worthwhile to remark that such as systems are not so uncommon in real examples, if the lagrangian coordinates are properly chosen.

\noindent
Lastly, it is worth to check the adjustement of (\ref{voronec}) when one or more than one of the constraints (necessarily of the form (\ref{constrlin})) is integrable, that is there  exists a function $f^{(\mu)}$ such that $\alpha_{\mu,i}(q_1,\dots, q_m)=\frac{\partial f^{(\mu)}}{\partial q_i}$ for some $\mu$ between $1$ and $k$ and for each $i=1,\dots, m$. If one defines
$$
T_\mu=T^*(q_1,\dots, q_m,\dots, q_{\mu-1}, f^{(r)}(q_1,\dots, q_\mu), q_{\mu+1}, \dots, q_n, {\dot q}_1, \dots, {\dot q}_\mu,t)
$$
which does not contain $q_\mu$, one has with respect to (\ref{voronec}) $\dfrac{d}{dt}\dfrac{\partial T^*}{\partial {\dot q}_i}-\dfrac{\partial T^*}{\partial q_i}	-\alpha_{\mu,i}\dfrac{\partial T^*}{\partial q_{m+\mu}}=
\dfrac{d}{dt}\dfrac{\partial T_r}{\partial {\dot q}_i}-\dfrac{\partial T_r}{\partial q_i}$ and $\beta_{ij}^\mu=0$, $i,j=1,\dots, m$.
Clearly, if any of the constraints (\ref{constrlin}) is integrable, the left side of (\ref{voronec}) reduces to the lagrangian binomial  $\frac{d}{dt}\frac{\partial {\hat T}}{\partial {\dot q}_i}-\frac{\partial {\hat T}}{\partial q_i}$ characteristic of holonomic systems, where 
$$
{\hat T}(q_1, \dots, q_m, {\dot q}_1, \dots, {\dot q}_m, t)=T^*(q_1, \dots, q_m, f^{(1)}(q_1, \dots, q_m), \dots, f^{(k)}(q_1, \dots, q_m), {\dot q}_1, \dots, {\dot q}_m, t).	
$$

\section{Comparing the two methods}

\noindent
For the purpose of writing the equations of motion we formally followed two different ways and we achieved them in the form (\ref{vnl2}) and in the form (\ref{vnl}): it is not so evident that they turn out to be the same set of equations.
A not eluding discrepancy is the presence of the second derivatives of the functions $\alpha_\nu$ in (\ref{b}), whereas they are not comparing in the coefficients (\ref{coeffcdef}). In order to verify that the nonlinear Voronec equations (\ref{vnl}) correspond to (\ref{vnl2}) we have to come back to (\ref{encin}) which gives, recalling (\ref{calq}), 
$T=\frac{1}{2}
\sum\limits_{i,j=1}^n g_{i,j}{\dot q}_i {\dot q}_j +\sum\limits_{i=1}^n b_i {\dot q}_i +c$ where $g_{i,j}$ are the same functions as in (\ref{gxi}) and $b_i(q_1,\dots, q_n,t)=\dfrac{\partial {\bf X}^{(\mathsf{M})}}{\partial q_i}\cdot \dfrac{\partial {\bf X}}{\partial t}$, $c(q_1,\dots, q_n,t)=\frac{1}{2}
\dfrac{\partial {\bf X}^{(\mathsf{M})}}{\partial t}\cdot 
\dfrac{\partial {\bf X}}{\partial t}$. By means of (\ref{trid}) we can write explicitly $T^*$ in the following way:
\begin{eqnarray}
\nonumber
T^*(q_1\dots, q_n, {\dot q}_1, \dots, {\dot q}_m,t)&=&
\frac{1}{2}\left( \sum\limits_{r,s}^m g_{r,s}{\dot q}_r {\dot q}_s+
\sum\limits_{\nu, \mu=1}^k g_{m+\nu,m+\nu} \alpha_\nu \alpha_\mu \right)+
\sum\limits_{r=1}^m \sum\limits_{\nu=1}^k g_{r,m+\nu}{\dot q}_r \alpha_\nu\\
&+&
\label{tmobile}
\sum\limits_{r=1}^m b_r {\dot q}_r +\sum\limits_{\nu=1}^k b_{m+\nu}\alpha_\nu+c
\end{eqnarray}
where $\alpha_\nu(q_1\dots, q_n, {\dot q}_1, \dots, {\dot q}_m,t)$ are always the functions (\ref{constrexpl}). Whenever the geometric constraints are stationary, that is ${\bf X}={\bf X}(q_1,\dots, q_n)$, the second line in (\ref{tmobile}) disappear.

\noindent
The rearrangement of the various terms in (\ref{vnl}) when $T^*$ assumes the just written form is regulated by the following
\begin{prop}
	Assume that equations (\ref{vnl}) are written choosing $T^*$ as in (\ref{tmobile}). Then, for each $i=1,\dots, m$ the summation $-\sum\limits_{\nu=1}^k \dfrac{\partial T}{\partial {\dot q}_{m+\nu}} B_{i}^\nu$ coincides with the sum of the three terms
	$$
	\begin{array}{ll}
	(i) &-\sum\limits_{r,s=1}^m \sum\limits_{\nu,\mu, \sigma=1}^k \left(g_{r,m+\nu}{\dot q}_r+g_{m+\nu, m+\mu} \alpha_\mu+b_{m+\nu}\right)
	\left( \dfrac{\partial^2 \alpha_\nu}{\partial {\dot q}_i \partial q_s}{\dot q}_s +
	\dfrac{\partial^2 \alpha_\nu}{\partial {\dot q}_i \partial {\dot q}_s}{\q2dot^{..}}_s+
	\dfrac{\partial^2 \alpha_\nu}{\partial {\dot q}_i \partial q_{m+\sigma}}\alpha_\sigma
	+\dfrac{\partial^2 \alpha_\nu}{\partial {\dot q}_i \partial t}\right),\\ 
	(ii) &  \sum\limits_{r=1}^m\sum\limits_{\nu=1}^k \left( g_{r,m+\nu}
	{\dot q}_r+g_{m+\nu,m+\mu}\alpha_\mu+b_{m+\nu}\right)\dfrac{\partial \alpha_\nu}{\partial q_i}, 	\\ 
	(iii) & \sum\limits_{r,s=1}^m \sum\limits_{\nu,\mu, \sigma=1}^k \left(g_{r,m+\nu}{\dot q}_r+g_{m+\nu, m+\mu}
	\alpha_\mu+b_{m+\nu}\right) 
	\dfrac{\partial \alpha_\sigma}{\partial {\dot q}_i}\dfrac{\partial \alpha_\nu}{\partial q_{m+\sigma}}.  
	\end{array}
	$$
In addition, the same terms with opposite sign appear respectively among the terms of $\dfrac{d}{dt}\left(\dfrac{\partial T^*}{\partial {\dot q}_i}\right)$ {\rm(}terms $(i)${\rm)}, 
	of $-\dfrac{\partial T^*}{\partial q_i}$ {\rm(}terms $(ii)${\rm)} and of
	$-\sum\limits_{\nu=1}^k\dfrac{\partial T^*}{\partial q_{m+\nu}}\dfrac{\partial \alpha_\nu}{\partial {\dot q_i}}$ {\rm(}terms $(iii)${\rm)}. 
	Therefore, all terms in $-\sum\limits_{\nu=1}^k \dfrac{\partial T}{\partial {\dot q}_{m+\nu}} B_{i}^\nu$ vanish and the remaining terms of (\ref{vnl}) coincide precisely with (\ref{vnl2}).
\end{prop}

\noindent
{\bf Proof.} The check is based on the following preliminary formulae 
$$
\begin{array}{l}
\dfrac{\partial T^*}{\partial {\dot q}_i} =
\sum\limits_{r=1}^m \sum\limits_{\nu,\mu=1}^k
\left( 
g_{i,r}{\dot q}_r+g_{i,\nu} \alpha_\nu + 
\left( 
g_{r,m+\nu}{\dot q}_r
+ g_{r,m+\nu} g_{m+\nu, m+\mu} \alpha_\mu +b_{m+\nu}
\right) 
\dfrac{\partial \alpha_\nu}{\partial {\dot q}_i} 
\right) +b_i \\
 i=1,\dots, m,  \\
 \vspace{.5truecm}
\dfrac{\partial T^*}{\partial q_i} =
\sum\limits_{r,s=1}^m  \sum\limits_{\nu,\mu=1}^k 
\left(
\frac{1}{2}
\dfrac{\partial g_{r,s}}{\partial q_i}{\dot q}_r{\dot q}_s+\frac{1}{2}
\dfrac{\partial g_{m+\nu,m+\mu}}{\partial q_i}\alpha_\nu \alpha_\mu +
\left(
\dfrac{\partial g_{r,m+\nu}}{\partial q_i}\alpha_\nu+
\dfrac{\partial b_r}{\partial q_i}\right){\dot q}_r \right. \\
\left. + \left( g_{m+\nu, m+\mu}\alpha_\mu + g_{r,m+\nu}{\dot q}_r+b_{m+\nu}\right)
\dfrac{\partial \alpha_\nu}{\partial q_i} \right)+\frac{1}{2}\dfrac{\partial c}{\partial q_i}
\quad i=1,\dots, n, \\
\left.\dfrac{\partial T}{{\dot q}_{m+\nu}}
\right\vert_{{\dot q}_{m+\mu}=\alpha_\mu, \mu=1,\dots, k}=
\sum\limits_{r=1}^mg_{m+\nu, r}{\dot q}_r +\sum\limits_{\mu=1}^k g_{m+\nu, m+\mu}\alpha_\mu +b_{m+\nu} \quad 
 \nu=1,\dots, k.
\end{array}
$$
The last equality, joined with (\ref{b}), gives immediately the expression of  
$-\sum\limits_{\nu=1}^k \dfrac{\partial T}{\partial {\dot q}_{m+\nu}}B_i^\nu$ as the sum of $(i)$, $(ii)$ and $(iii)$. 
On the other hand, the calculus of $\dfrac{\partial T^*}{\partial q_i}$ considered either in the case $i=1,\dots,m$ and in the case $i=m+1,\dots, n$ easily exhibit the terms $(ii)$ and $(iii)$ among the terms of $\dfrac{\partial T^*}{\partial q_i}$  and of $\sum\limits_{\nu=1}^k\dfrac{\partial T^*}{\partial q_{m+\nu}}\dfrac{\partial \alpha_\nu}{\partial {\dot q_i}}$, in the two respective cases. Owing to the minus sign, they mutually cancel.
Furthermore, one promtly realizes that the explicit calculus of $\dfrac{d}{dt}\dfrac{\partial T^*}{\partial {\dot q}_i}$ contains also the opposite of $(i)$ among the terms, so that $-\sum\limits_{\nu=1}^k \dfrac{\partial T}{\partial {\dot q}_{m+\nu}} B_{i}^\nu$ entirely vanishes.
If we carry out the calculus by replacing the formulae written at the beginning of the proof in (\ref{vnl}), it is straightforward to check the matching of (\ref{vnl}) with (\ref{vnl2}). $\quad \square$

\section{Conclusions}

\noindent
The performed analysis embraces a dual aim: to present a formulation for nonlinear kinematical constraints and to extend the classical Voronec equations for such systems.

\noindent
The approach we followed concerning systems with nonholonomic constraints is based on identifying the set of admitted velocities with (\ref{subspace2}), which is ispired to the linear situation (\ref{subspace}).
In that way, the concept of ideal constraints makes sense: the calculation of the equations of motion starting from the D'Alembert principle is straightforward and it ends with the explicit equations (\ref{vnl2}).

\noindent
Changing the perspective, if one adheres the new context of nonlinearity to the classical procedure of Voronec and 
$\check{\rm C}$aplygin performed in the linear case, the formal derivation of the equations via the Lagrangian function leads to (\ref{vnl}), where the Lagrangian structure of the equations of motion is evident and the contribute of   nonlinearity (\ref{constrexpl}) with respect to the linear (classic) cases (\ref{voronec}) and (\ref{capligin}) is understable. 
At the same time, in (\ref{vnl2}) and in (\ref{vnl}) both the integer constraints and the kinematical constraints can  depend explicitly on time: once again, the additional terms due to the mobility of the constraints are clear in the coefficients (\ref{coeffcdef}) and (\ref{b}).

\noindent
The formulation (\ref{vnl2}) has the advantage of exhibiting directly the explicit form of the equations of motion and corresponds  to the calculus via the acceleration energy of Appell. The second method (\ref{vnl}) prioritizes the preservation of the Lagrangian-type equations and gives the extension to the nonlinear case of the Voronec equations. 

\noindent
In both cases no quasi--coordinates nor quasi--velocities are introduced, in step with the Voronec's approach: if, on one side, the use of such coordinates may support the analytical procedure, on the other side it is generally not evident how to define them in a properly way; moreover, the calculation of the equations is simplified since no inverse matrices (required by Hamel--Boltzmann type methods) are necessary.

\noindent
The apparent disparity of the two types of equations of motions (\ref{vnl2}) and (\ref{vnl}) motivates the comparison performed in Section 4, for a quite general system with kinetic energy as in (\ref{tmobile}): with regards to it, the result of Proposition 4.1 points out that in the Voronec--type of motion of  several terms are redundant. 
Even in the linear case, this sort of check seemeed us to be missing in literature.

\noindent
Nonholonomic constraints are undoubtely investigated mainly in the linear case: we mention \cite{gant}, \cite{lurie}, \cite{pars} among the important texts on Mechanics dwelling almost solely on the linear situation. 
By the other hand, we refer to \cite{papastr} for a comprehensive list of references abuot nonlinear nonholonomic systems, where the scarcity of literature - especially in english - on such an important topic is underlined. 

\noindent
On the other hand, greater attention is devoted in literature to the employment of quasi--coordinates and pseudo--velocities, so that dynamical equations based on the explicit step (\ref{constrexpl}) 
are more infrequent.
As a matter of fact, mainly in the nonlinear case it may happen not to be so worthwhile to rearrange the generalized velocities in order to introduce 
pseudo--velocities, especially whether there is no evidence of simplifying the mathematical problem.
The method adopted here for nonlinear constraints reveals the set of permitted velocities, which is the key point for 
``projecting'' the Newton's law, by an elementary way and writing the equations of motion is straightforward.

\noindent
An  interesting point  connected to our approach consists in dealing with higher order constraints equations, which can be  sketched by the equations ${ \Phi}_j
(q_1, \dots, q_n, {\dot q}_1, \dots, {\dot q}_n,$ 
$\dots,  {\q2dot^{p \cdot}}_1, \dots,  {\q2dot^{p \cdot}}_n, t)$, 
where $p \cdot$ stands for the sequence of $p$ dots, $p\geq 2$: 
having in mind the lagrangian property 
$\dfrac{\partial \bfx2dot^{\kappa \cdot}}{\partial {\q2dot^{\kappa \cdot}}_i}=\dfrac{\partial {\bf X}}{\partial q_i}$, $\kappa=1,2,\dots$ one enables to extend (\ref{subspace2}) to higher derivatives, 
so that the method can be generalized in a natural way to higher-order constraint equations.
Both the procedures we presented, leading up to (\ref{vnl2}) the one and to (\ref{vnl}) the second, are once again 
accessible.

'\end{document}